\documentclass[10pt]{article}
\usepackage{nicefrac}

\usepackage{geometry}
\usepackage{hyperref}
\usepackage{graphicx}
\usepackage{amsmath, amssymb}
\usepackage{natbib}

\title {A Local Probe Mass Spectrometer for Localized and Sensitive Product Detection in Environmental Electron Microscopy} 
\author{
  Saleh Firoozabadi $^{1}$ \\
  Timofei Ivanov $^{1}$\\
  Frederik Stender $^{1}$\\
  Julian Grahl $^{2}$\\
  Stephan Schulz $^{2}$\\
  Christian Jooss $^{1}$\\
  Tobias Meyer $^{1,*}$\\
  \\
  \parbox{0.8\textwidth}{\centering
    \textit{1. Institute of Materials Physics, University of Göttingen, Friedrich-Hund-Platz 1, Göttingen, 37077, Lower Saxony, Germany} \\
    \textit{2. Institute of Inorganic Chemistry, University of Duisburg-Essen, Universitätsstraße 5, Essen, D-45141, North Rhine-Westphalia, Germany} \\
    \textit{*Corresponding author: tmeyer@uni-goettingen.de}\\
    \textit{Saleh Firoozabadi and Timofei Ivanov contributed equally to this work.}\\
  }
}

\begin{document}

\maketitle
\begin{abstract}
\noindent
Aberration-corrected environmental transmission electron microscopy (ETEM) enables atomic-resolution imaging of dynamic catalytic processes. Correlating atomic-scale structural changes with reaction products detected by mass spectrometry offers a powerful route to uncover catalytic mechanisms. However, current approaches face fundamental limitations: closed-cell ETEM setups suffer from diffuse scattering by SiN windows, degrading spatial resolution and sensitivity, while open-cell configurations enable high-resolution imaging and maintain high sensitivity but suffer from significant dilution of reaction products during transport to the mass spectrometer (MS). To overcome these challenges, we develop a Local Probe Mass Spectrometer (LPMS) integrated with aberration-corrected ETEM. The setup combines a DENSsolution Stream holder with a MS. To preserve spatial resolution, both top and bottom SiN membranes of the MEMS chip are removed, while the gas environment is maintained via the ETEM chamber. Reaction products are sampled locally via a micro-capillary positioned near the catalyst and connected to a holder gas line that delivers the gas to the MS. Initial validation in environmental SEM confirmed controlled gas delivery to the MS. Co$_3$O$_4$ nanoplates serve as a model catalyst due to their inherent electron transparency, enabling atomic-resolution imaging without FIB lamella preparation and associated ion-beam damage. A novel micro-shuttle transfer strategy enables controlled placement of a defined number of nanoplates at the reaction site with precise crystallographic orientation. This establishes the foundation for quantitative structure–reactivity correlation by enabling simultaneous, spatially resolved detection of reaction products and atomic-scale structural dynamics.
\end{abstract}

\textbf{keywords:} Local Probe Mass Spectrometry, Catalyst transfer, Gas-phase electrochemistry, Environmental EM, Aberration-corrected TEM

\section{Introduction}\label{intro}
Hydrogen production is crucial for renewable energy storage due to its high energy density, non-toxic nature, and environmental benefits \citet{sarmah2023sustainable}. In particular, producing hydrogen from water is attractive because water is an abundant and readily available resource. However, the sluggish oxygen evolution reaction (OER) at the anode, driven by multi-electron transfer processes and Gibbs free energy differences between oxidation steps, remains a key challenge \citet{krishtalik1986energetics}. The ideal free-energy change for each OER step is 1.23 eV; however, a higher potential barrier in at least one intermediate step leads to an overpotential. The lowest possible overpotential can be theoretically estimated using density functional theory. Assuming a four-electron transfer mechanism and a single active site results in a linear scaling relation \citet{rossmeisl2007electrolysis}, which is applicable to a wide variety of oxide catalyst materials \citet{man2011universality}.

Nevertheless, there are catalyst systems that exhibit lower overpotentials than predicted by the scaling relations reported in  \citet{rossmeisl2007electrolysis,man2011universality}. The underlying mechanism responsible for this behavior remains under debate, with several explanations proposed in the literature. One commonly discussed assumption involves a dual-site reaction mechanism \citet{conway1964electrochemical,song2019unconventional,gono2020oxygen}. 

Another approximation used in the calculations of \citet{rossmeisl2007electrolysis} is the assumption of a frozen electrocatalyst surface, thus neglecting structural rearrangements of surface atoms during the reaction. Therefore, resolving the atomic-scale  structural dynamics of the electrocatalyst-water interface during the OER can provide deeper insight into the underlying mechanisms. Aberration-corrected ETEM enables real-space atomic-resolution studies of catalysts under reactive conditions, whether in liquid or gas phases. For gas-phase environments, this includes near-industrial pressures in closed-cell configurations or several millibars in differentially pumped systems.

In liquid-phase, structural and analytical resolution is often limited by electron scattering from the liquid layer and window membranes \citet{hwang2020situ}. Improvements have been achieved by reducing the liquid thickness, for example to ~30 nm via in-situ bubble formation during water splitting \citet{serra2021nanoscale}, or by encapsulating liquids between graphene layers to achieve atomic resolution \citet{yuk2012high}. Closed-cell gas-phase concepts for in-situ TEM were first proposed by \citet{heide1962electron} in 1962. The initial design employed two specimen grids separated by a thin metallic foil at a defined distance and sealed within the specimen holder. Subsequent MEMS-based developments enabled atomic-scale imaging at pressures up to 1.2 bar and under heating conditions \citet{creemer2008atomic,creemer2010mems}. 

Closed-cell ETEM typically relies on SiN window membranes, which introduce diffuse scattering and limit both spatial resolution as well as sensitivity to signals from few-atom species, whereas open-cell ETEM avoids window scattering and enables high-resolution imaging \citet{jinschek2012image}. Such open-cell and differentially pumped ETEM configurations enable in-situ studies of dynamic chemical reactions at the atomic scale \citet{boyes1997environmental,hansen2001atomic,helveg2004atomic}. For OER specifically, the emergence of a ~0.6 nm dynamic layer at the electrocatalyst surface has been reported \citet{ronge2021atom}, and direct visualization of surface atom dynamics at the electrocatalyst surface under reaction conditions has been demonstrated \citet{lole2020dynamic}. Taken together, these observations indicate that electrocatalyst surfaces undergo dynamic structural changes under reaction conditions, thereby challenging the validity of the frozen surface approximation commonly used in theoretical models suggested by \citet{rossmeisl2007electrolysis}. 

Despite open-cell ETEM allowing to track motion of individual atoms with high sensitivity, there are critical differences between the electrochmical evironment in ETEM compared to the in-operando electrochemical conditions. To preserve the high vacuum in the rest of the TEM column, the pressure in the ETEM specimen environment is limited to a few mbar \citet{BIRAN2026114328}. Consequently, the accessible pressure remains much lower than in realistic electrocatalytic operating conditions. Furthermore, in contrast to electrochemical systems, pH control in water vapor has not yet been realized. Thus, it is unclear till date if the observed atomic dynamics are actually related to the mechanism occuring during electrocatalytic water splitting. In addition to these environmental differences, the electron beam itself can modify the local chemical environment. Since electron-beam-induced plasma formation has been reported \citet{LINDNER2023113629}, it is important to understand the chemical environment experienced by the catalyst, which may locally differ from pure water vapor. Under these conditions, identifying the formed reaction products is critical to determine whether the catalyst is observed in an active state.

To this end, previous studies have employed sacrificial agents that induce solid-phase growth at active sites, providing indirect evidence of O$_2$ evolution \citet{raabe2012situ,mildner2015environmental}. Alternatively, integration of ETEM with mass spectrometry has been demonstrated for both closed-cell \citet{vendelbo2014visualization,plodinec2020versatile} and open-cell \citet{raabe2012situ,muto2019environmental} configurations. While these approaches represent important steps toward correlating structural dynamics with reaction products, open-cell ETEM in particular suffers from significant dilution of reaction products as gases travel from the ETEM chamber to the mass spectrometer (MS). This arises from the small volume of the reaction site in ETEM catalysis experiments, the large volume of the sample chamber, and the extended transport path from the reaction site to the MS, leading to substantial dilution before detection. 

In this study, we develop a local probe mass spectrometer (LPMS) for direct correlation of the active catalyst state with reaction products under in-situ conditions. The approach integrates aberration-corrected ETEM with a quadrupole mass spectrometer, enabling simultaneous atomic-resolution imaging and sensitive product detection. As a model catalyst system, we employed Co$_3$O$_4$ nanoplates exhibiting a low overpotential reflecting a possibility of breaking the scaling relations between the binding energies of OER intermediates. Furthermore, their inherent electron transparency enables atomic-scale characterization without the need for lamella preparation, thereby avoiding ion-beam damage. In addition, a dedicated transfer approach is developed to position a defined number of nanoplates at the reaction site with controlled crystallographic orientation. This provides a route toward quantitative structure–reactivity correlation under in-situ conditions.

\section{Integrated ETEM–LPMS Platform}\label{method}
\subsection{System Design and General Assembly}
\begin{figure}[htbp]
  \centering
   \includegraphics[width=0.5\linewidth]{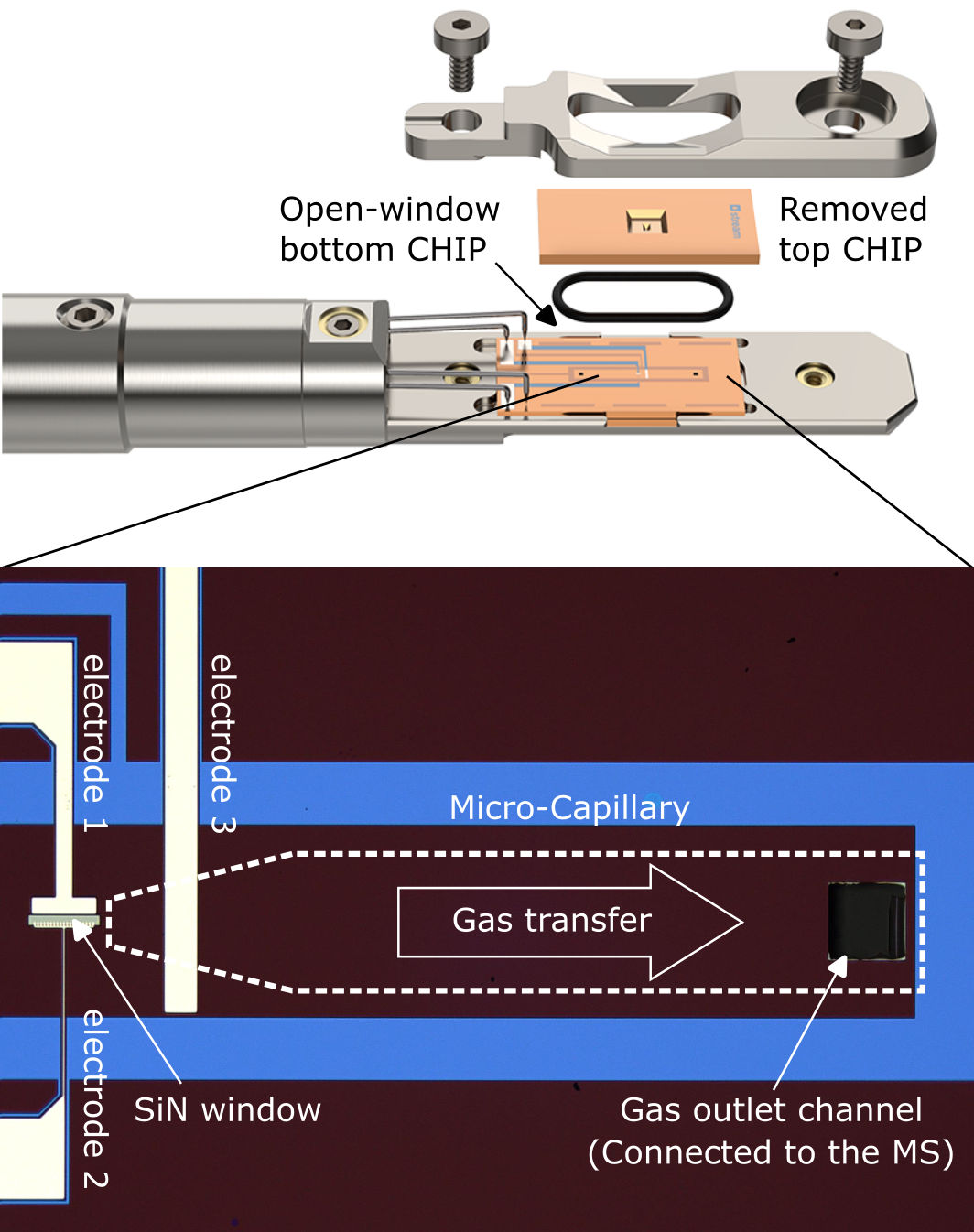}
    \caption{Overview of the LPMS setup. The system integrates a DENSsolutions Stream holder equipped with a micro-capillary for localized gas sampling. Reaction products generated at the sample are transferred through the capillary to the mass spectrometer via the holder’s gas line. The holder illustration is adapted in part from the manufacturer-provided schematic of the DENSsolutions Stream holder \citet{dens_stream}.}\label{assembly}
\end{figure}

The LPMS experimental setup combines a quadrupole MS (MKS V2000P) with a MEMS based in-situ holder (Stream, DENSsolutions). As illustrated in Fig. \ref{assembly}, a micro-capillary positioned a few tens of micrometers from the sample site transfers gas products to one of the TEM holder’s gas outlet channels, which is connected to the MS.  The MEMS chip contains three electrodes, which are used as follows: Electrode 1 and 2 are connected both to a shuttle on which the nanoparticles are placed (see section \ref{sec:sampleTrans}) hence the electrical connection to the sample can be checked by measureing the resistance between them. Electrode 3 is connected to the capillary allowing to use it as a counter electrode as well as to attract and repell ionic reaction intermediates and products.

The reaction gas environment is provided either by an open-cell ETEM (FEI Titan 80–300 G2 ETEM) or, for initial testing, an environmental SEM (FEI Quanta 200). 
In the ETEM configuration, a spatial resolution of approximately an ångström  is achieved using a CEOS CETCOR image corrector. To preserve this spatial resolution, the top chip of the Stream holder is removed, and a hole is milled in the bottom chip SiN window.

\subsection{Local Probe Capillary}
\begin{figure*}[htbp]
  \centering
   \includegraphics[width=1.0\linewidth]{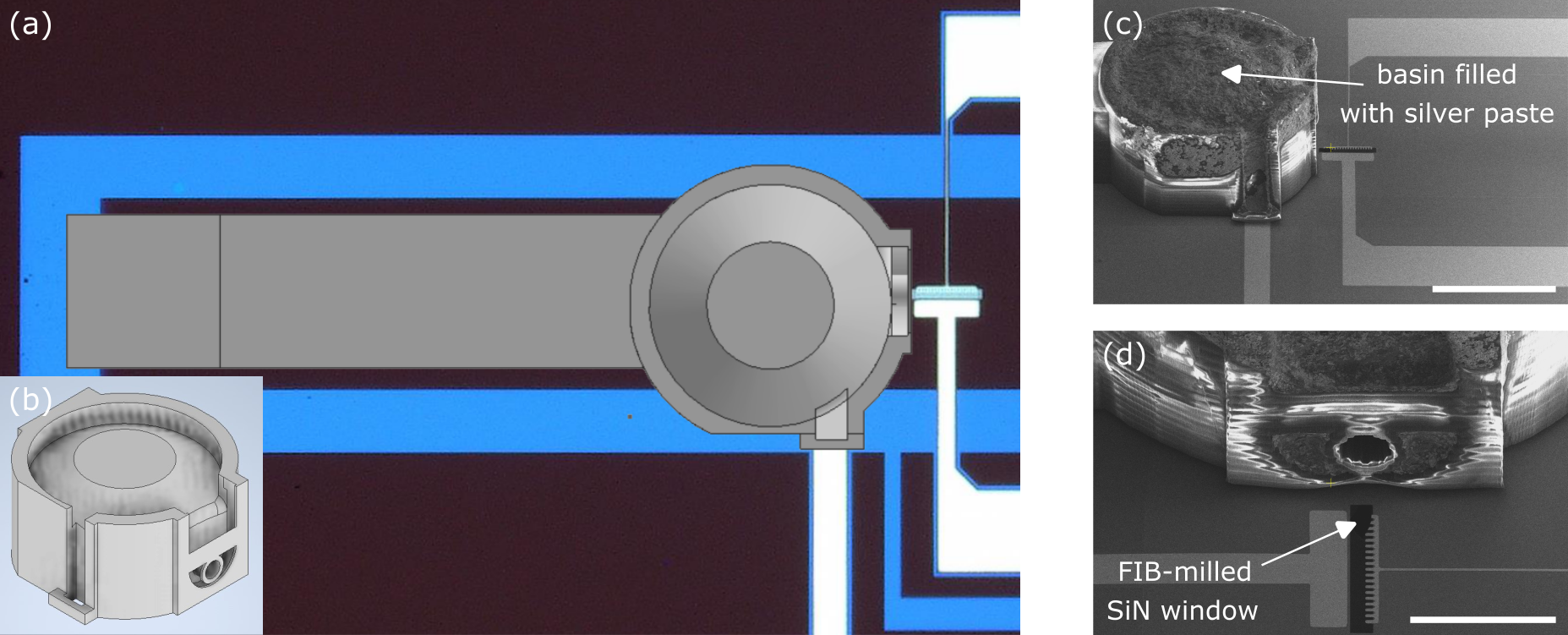}
    \caption{Geometry, positioning, and electrical design of the micro-capillary on the MEMS chip. \textbf{a} Overview of the full capillary structure integrated on the chip. \textbf{b} Capillary head geometry enabling electrical connection to the electrode. \textbf{c–d} Detailed views of the capillary head from different orientations, highlighting its geometry and alignment relative to the chip window and electrodes. Scale bars: (c) 500 \(\mu\)m, and (d) = 200 \(\mu\)m}\label{cap}
\end{figure*}

The local probe capillary is fabricated using two-photon polymerization, a technique that enables the fabrication of high-resolution, three-dimensional micrometer-scale structures using pulsed lasers (Nanoscribe Photonic Professional GT) \citet{maruo1997three}. Figure \ref{cap}a shows the positioning of the capillary between the SiN window and the holder’s gas outlet channel, crossing electrode 3. A detailed view of the capillary head geometry is presented in Fig. \ref{cap}b. The capillary has an entrance diameter of 80 $\mu$$m$, located at a height of 90 \(\mu\)m above the chip surface, while the inner tube diameter is 240 \(\mu\)m.

To enable electrical conduction at the capillary head, the basin-shaped geometry shown in Fig. \ref{cap}b is employed. The basin is filled with silver paste, establishing electrical contact with electrode 3. Connecting a Au wire (20um diameter) to the front of the capillary and measuring the resistance to electrode 3 yielded a value of 130 $\Omega$ proving the possibility to use the front face of the capillary as a counter electrode. SEM images in Fig. \ref{cap}c–d show a fully fabricated capillary on the MEMS chip, including the silver-filled head, positioned in close proximity to the open window region where the catalyst is located and the reaction occurs, and electrically connected to electrode 3. To ensure a leak-tight connection, a two-component epoxy resin is applied around the outlet junction of the printed capillary, where it wets the surrounding side walls and seals the capillary–channel interface. A detailed analysis of the influence of the capillary entrance diameter on the flow resistance is presented in section \ref{sec:GasFlow}.

\subsection{Gas Transport and Flow Characterization}\label{sec:GasFlow}
A series of finite element simulations were performed using the Computational Fluid Dynamics Module in COMSOL Multiphysics\textsuperscript{\textregistered} \citet{comsol54}, followed by a experimental testing in a rough vacuum chamber to analyze and characterize the flow through the micrometer-scale capillary.

\subsubsection{Gas Flow simulations}\label{sec:simFlow}
The gas flow is modelled assuming laminar flow. In the proposed setup, the overall flow rate is determined by the combined resistance of two tubes connected in series: (i) a tube connecting the mass spectrometer to the center of the TEM holder, with a length of 30 cm and a diameter of 150 \(\mu\)m, and (ii) the micro-capillary, with a variable entrance diameter, a length of 2830 \(\mu\)m, and an inner diameter of 240 \(\mu\)m. The inlet pressure is defined by the ETEM environment and typically ranges from 1 to 3 mbar, while the outlet pressure at the mass spectrometer side is assumed to be $10^{-2}$ mbar in the simulations, corresponding to the pressure at the entrance slit of the mass spectrometer. The entrance length is set to 72 \(\mu\)m.

Based on this model, the influence of the capillary entrance diameter on the flow rate was evaluated. Fig. \ref{flow-sim-exp}a shows the simulated flow rate as a function of capillary entrance diameter for two ETEM pressures of 1 and 3 mbar. The results show a significant decrease in flow rate for capillary diameters below 50 \(\mu\)m, whereas the values remain relatively constant for larger diameters. Thus, for entrance apertures larger than 50 \(\mu\)m the flow rate is determined by the flow channel of the holder rather than the micro-capillary.

\subsubsection{Experimental flow resistance characterization}\label{sec:expFlow}
The leak rate is determined by relating the pressure increase inside a previously evacuated container to the gas inflow. The experimental setup consists of an aluminum container with a volume of $V_c = 0.023\,\mathrm{m^3}$, equipped with an outlet for evacuation and an inlet connected to the holder. After evacuation, a Pirani Compact gauge monitors the pressure increase $p_c(t)$ caused by gas entering through the holder. Here, $p_{in}$ denotes the ambient pressure and $p_c \ll p_{in}$ at early times. By selectively sealing the two gas channels of the holder using plugs, different flow resistances can be determined and unintended leak paths identified.

Acknowledging the large container volume, the inflowing gas does not significantly affect the pressure difference at relevant time scales, resulting in a quasi-stationary flow with $p_{in} \gg p_c$ and an approximately constant inflow rate. Thus, the pressure increase is linear in time, and the flow resistance $R_F$ can be determined from the slope $\frac{\partial p_c}{\partial t}$, the known volume $V_c$, and the inlet pressure $p_{in}$: 
\begin{equation}
   \dot{p_c} V_c = \frac{1}{R_F} \cdot (p_{in}^2 -p_c^2) \approx \frac{1}{R_F} \, p_{in}^2  . \label{formul2}
\end{equation}

Additionally the flow resistance can also be calculated using the Hagen-Poiseuille-Equation, yielding: 

\begin{equation}
   R = \frac{16 \mu L}{\pi r^4}.
\end{equation}\label{formula1}

\begin{figure*}[htbp]
  \centering
   \includegraphics[width=1.0\linewidth]{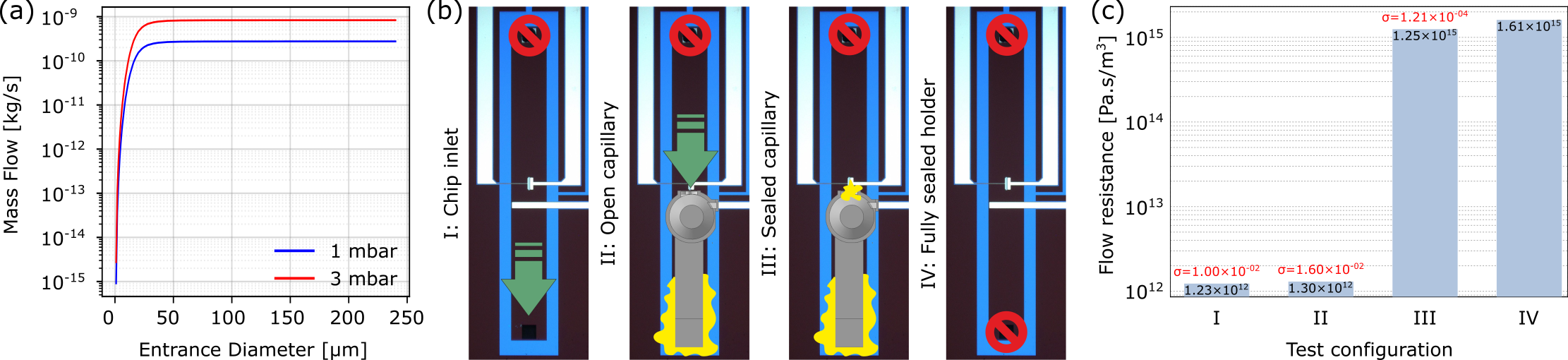}
    \caption{Gas transport and flow characterization of the capillary. \textbf{a} Simulated dependence of flow rate on capillary entrance diameter. \textbf{b} Different experimental configurations used to characterize gas flow through the capillary: Configuration I, venting through the chip inlet; Configuration II, venting through the glued, open capillary; Configuration III, venting through the glued, sealed capillary; and Configuration IV, holder fully sealed with plugs. \textbf{c} Comparison of the measured flow resistance for the different configurations, demonstrating that the capillary connection to the chip outlet is leak-tight and preserves the overall flow rate.}\label{flow-sim-exp}
\end{figure*}

To evaluate the leak-tightness of the capillary-chip connection and verify that the capillary does not significantly impede the flow rate, four leak-test configurations were employed (Fig. \ref{flow-sim-exp}b). Panels (a) and (b) compare the flow through a standard chip channel with that of a holder containing a printed capillary. The results demonstrate a negligible decrease in flow rate, confirming that the capillary's presence does not restrict the path. Furthermore, a comparison between the front-sealed capillary (Fig. \ref{flow-sim-exp}c) and the plug-blocked holder (Fig. \ref{flow-sim-exp}d) indicates that the interface between the capillary and the chip outlet is sufficiently leak-tight. This suggests that the observed leak rate is primarily determined by the holder setup, while any additional leakage arising from our modifications is negligible.

Using the mass flow rate obtained from the simulations in Fig.~\ref{flow-sim-exp}a for an inlet diameter of 80~\(\mu\)m, the flow resistance was calculated as $4.41 \cdot 10^{11}\,\nicefrac{Pa\cdot s}{m^3}$. Applying Eq.~\ref{formula1} to the connecting tube, assuming that the capillary contributes only marginally to the overall flow resistance, yields a value of $8.78 \cdot 10^{11}\,\nicefrac{Pa\cdot s}{m^3}$ for a tube length of 30 cm and a diameter of 150~\(\mu\)m. The analytical estimate is thus approximately a factor of two higher than the simulation-based value. Both values are, however, consistent with the experimental data at the order-of-magnitude level. The remaining deviation may be attributed to additional effects such as kinks, bending, and surface roughness, which are not included in either the simulations or the analytical estimate. In addition, at the applied pressure difference, the assumption of a linear pressure drop underlying the Hagen–Poiseuille relation may no longer be fully valid.

\section{Controlled Catalyst Integration: Micro-Shuttle Transfer Strategy}\label{sec:sampleTrans}
\subsection{Motivation} 
\begin{figure*}[htbp]
  \centering
   \includegraphics[width=1.0\linewidth]{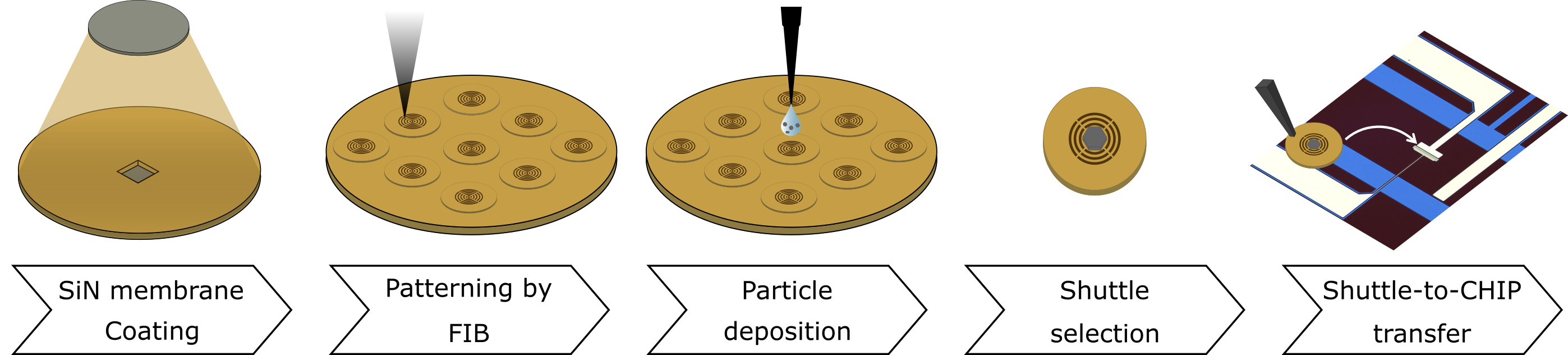}
    \caption{Schematic illustration of the micro-shuttle-based nanoparticle transfer workflow. The process consists of (i) Gold coating of a SiN membrane substrate, (ii) FIB patterning of micro-shuttles with a central window, (iii) deposition of nanoparticles via drop-casting, (iv) selection of individual shuttles with suitable particle loading, and (v) transfer of the selected micro-shuttle onto a MEMS chip.}\label{Transfer_Schematic}
\end{figure*}

Choosing Co$_3$O$_4$ nanoplates as a model catalyst system provides inherent electron transparency, thereby eliminating the need for FIB lamella preparation and avoiding associated ion-beam damage or electron-beam-induced contamination. The hexagonal plate morphology naturally corresponds to the [111] zone axis when the particles are oriented flat, which facilitates alignment of the crystal along this direction. In addition, the edges of the hexagonal plates expose \{220\} surface terminations, which have been reported to exhibit the second-highest OER activity  after \{222\} \citet{saddeler2020effect}. Since the \{222\} planes lie parallel to the plate surface, they are oriented perpendicular to the electron beam direction, making in-plane imaging in TEM practically impossible due to inherent geometric constraints. 
We propose a transfer method for the nanoplates that avoids uncontrolled cluster transfer while enabling control over their crystallographic orientation. Fig. \ref{Transfer_Schematic} illustrates the steps of this transfer approach, which involves drop-casting nanoplates onto a micrometer-scale conductive gold film with open windows, referred to as a “micro-shuttle”, followed by transfer to the MEMS chip without ion- or electron-beam exposure.

\subsection{Micro-Shuttle Fabrication and Patterning}\label{shuttle}

Silicon nitride membranes with a thickness of 100 nm and dimensions of 0.5 \(\times\) 0.5 mm were coated with approximately 2.5 \(\mu\)m of gold by thermal evaporation. An array of 7 \(\times\) 7 circular micro-shuttles, each with a diameter of 60 \(\mu\)m and a central window, was then patterned on the coated grid using FIB milling. An SEM micrograph of the resulting gold film, with a magnified view of a single shuttle shown as an inset, is presented in Fig.~\ref{transfer_SEM_TEM}a.

\subsection{Catalyst Deposition, On-Shuttle Calcination}

The nanoparticles are distributed over the micro-shuttle-patterned gold film by drop-casting. Particle agglomeration can lead to misorientation and uncontrolled cluster transfer; therefore, it is essential to achieve a homogeneous dispersion on the gold film with ideally well-separated individual particles. To optimize the deposition procedure, the suspension was diluted in several steps and different solvent mixtures were evaluated. Improved particle dispersion was achieved by ultrasonication for several minutes; however, care was taken to avoid excessive sonication times that could damage the particles. The optimized deposition condition was obtained by diluting the particles to a concentration of 5~\(\mu\)g mL\(^{-1}\) in a water–ethanol mixture (3:1 volume ratio). The solution was ultrasonicated for 10 min, and 4~\(\mu\)L was drop-cast onto the micro-shuttles. To avoid electron-beam exposure and possible contamination of the deposited nanoparticles, SEM imaging after drop-casting was performed only in shuttle-free regions of the grid, as shown in Fig.~\ref{transfer_SEM_TEM}b. The image demonstrates a homogeneous dispersion of nearly individual particles across the grid.

The Co$_3$O$_4$ particles investigated in this study were synthesized via a solvothermal approach. Briefly, an aqueous solution of Co(NO$_3$)$_2$ was treated with a solution of oleylamine (OLA) in ethanol. The resulting Co(OH)$_2$ suspension was heated in an autoclave at 180 $^\circ$C for 15 h to produce hexagonal Co(OH)$_2$ platelets, which were subsequently transformed into the Co$_3$O$_4$ spinel phase by calcination in air at 300~$^\circ$C. To achieve a more homogeneous particle dispersion on the micro-shuttle film, we make use of the OLA capping agent present during the synthesis of the Co(OH)$_2$ particles. The uncalcined particles are first deposited onto the shuttle grid, which is thermally stable and can subsequently be heated in an oven to convert the particles to Co$_3$O$_4$. This on-grid calcination strategy enables improved particle separation prior to phase transformation. The calcination process is illustrated in Fig.~\ref{transfer_SEM_TEM}c--f. Fig. \ref{transfer_SEM_TEM}c and Fig. \ref{transfer_SEM_TEM}d show the successful deposition of a limited number of particles (here two) on a selected micro-shuttle, highlighted by dashed white lines for clarity. The corresponding selected area electron diffraction (SAED) patterns, acquired using a 150~\(\mu\)m aperture before (Fig.~\ref{transfer_SEM_TEM}e) and after (Fig.~\ref{transfer_SEM_TEM}f) calcination, confirm the sucessful phase transformation from Co(OH)$_2$ to Co$_3$O$_4$.

\subsection{FIB-Assisted Transfer and Electrical Validation}

\begin{figure*}[htbp]
  \centering
   \includegraphics[width=1.0\linewidth]{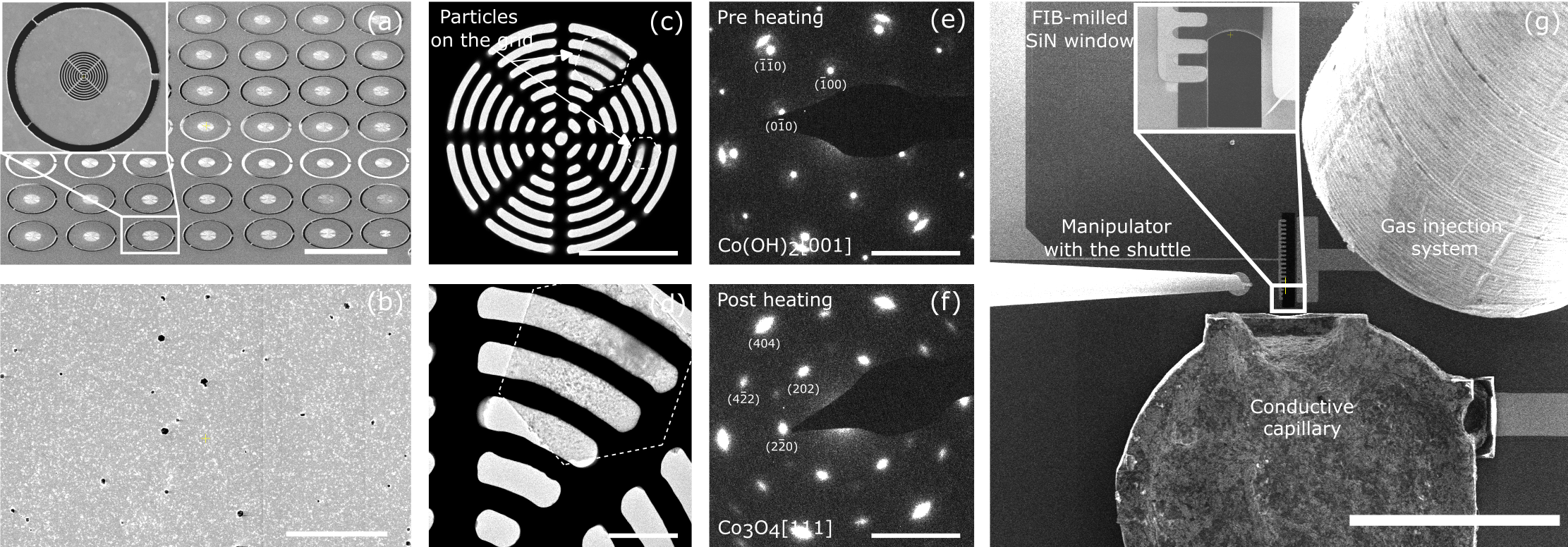}
    \caption{Controlled catalyst transfer using the micro-shuttle strategy. \textbf{a} FIB fabrication of micro-shuttles featuring a central window. \textbf{b} Optimization of drop-casting of Co$_3$O$_4$ nanoplates onto a grid patterned with micro-shuttles. SEM imaging is performed in shuttle-free regions of the grid after drop-casting to avoid electron-beam exposure and contamination of the deposited nanoparticles. \textbf{c,d} TEM images showing successful deposition of a limited number of particles (here two) on a single micro-shuttle; the particles are outlined by dashed white lines for clarity. \textbf{e,f} SAED patterns before and after calcination, confirming the phase transformation from Co(OH)$_2$ to Co$_3$O$_4$. \textbf{g} FIB-assisted transfer of a selected micro-shuttle onto the MEMS chip. Scale bars: (a) 100~$\mu$m, (b) 50~$\mu$m, (c) 5~$\mu$m, (d) 1~$\mu$m, (e,f) 5~nm$^{-1}$, and (g) 500~$\mu$m.
    }\label{transfer_SEM_TEM}
\end{figure*}

A micro-shuttle with the desired number and crystallographic orientation of nanoparticles is selected and transferred onto the MEMS chip using a micromanipulator inside the FIB system. As shown in the SEM micrograph in Fig.~\ref{transfer_SEM_TEM}g, the micromanipulator is attached to the shuttle in a manner that prevents the shuttle surface from being exposed to either the electron or ion beam. This approach also minimizes platinum redeposition during the welding step used to attach the shuttle to the MEMS chip.

Finally, the electrical conductivity of the chip was measured between the three holder electrodes to verify that no short circuit was introduced by material redeposition between electrode 3 and the other two electrodes, and to confirm reliable electrical contact between the micro-shuttle and electrodes 1 and 2. The resistance between electrode 3 and the other two electrodes exceeds 5 M\(\Omega\), while the resistance between electrodes 1 and 2, measured through the micro-shuttle, is 760 \(\Omega\). 
This large resistance difference, three orders of magnitude, ensures that micro-capillary can function as a stable counter electrode, while the low resistance path through the micro-shuttle enables reliable application of an external bias to the catalyst (working electrode). 

\section{Experimental Validation in ESEM}

\begin{figure*}[htbp]
  \centering
   \includegraphics[width=1.0\linewidth]{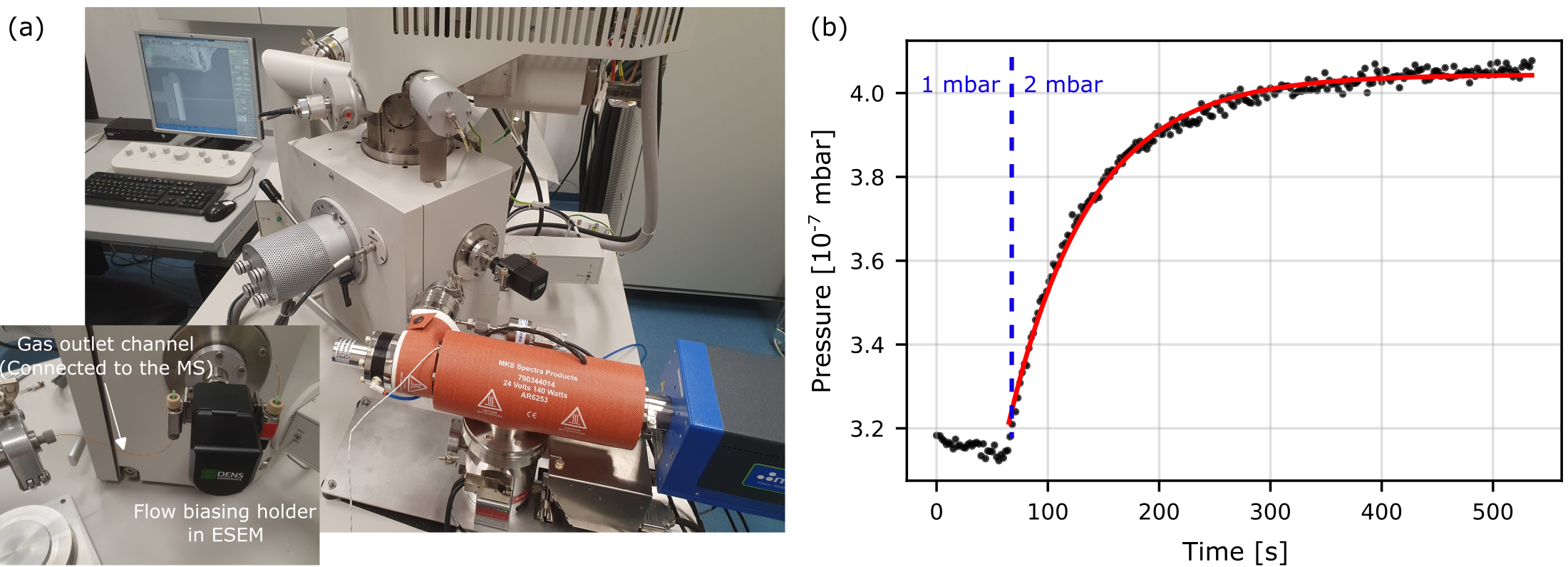}
    \caption{LPMS validation in an ESEM and temporal response characterization. \textbf{a} Validation of the LPMS setup in an ESEM using a custom-fabricated flange for TEM holder insertion. \textbf{b} Temporal evolution of the partial pressure of water vapor (mass 18 amu) measured by mass spectrometry following a step increase in chamber pressure from 1 to 2 mbar, together with an exponential fit used to extract the characteristic delay time constant.}\label{ESEM}
\end{figure*}

Initial validation of the LPMS setup was performed using an ESEM, which enables the introduction of water vapor at pressures of up to 10 mbar. Fig. \ref{ESEM}a shows the experimental setup, including the ESEM, the mass spectrometer, and a custom-fabricated flange that allows insertion of the TEM holder into the ESEM chamber.

This initial study was carried out to quantify the time delay between a step increase in ESEM chamber pressure from 1 to 2 mbar and the corresponding MS response, caused by the flow resistance of the micro-capillary and the holder’s glass tubing. The resulting temporal evolution of the water partial pressure at the mass-to-charge ratio channel of 18 amu is shown in Fig.~\ref{ESEM}b. An exponential fit was used to determine a characteristic delay time of 74 s. Thus, the response time is sufficiently short for correlation of structural dynamics with the reaction-product detection.

\section{Conclusion and outlook}

In this work, we developed and experimentally validated a local probe mass spectrometry (LPMS) approach for integration with environmental electron microscopy. Several key technical validations demonstrate the feasibility of the setup. First, the capillary connection to the chip outlet was shown to be leak-tight. Second, despite the small capillary entrance diameter required for localized product detection, the overall flow rate is preserved, confirming sufficiently low flow resistance. Third, a controlled micro-shuttle-based transfer strategy was established for the integration of electron-transparent, highly active catalyst particles onto the MEMS chip without exposing the sample to electron or ion beam irradiation during FIB transfer. Fourth, Electrical validation confirmed two key functionalities: (1) reliable application of an external potential to the catalyst via the conductive gold micro-shuttle, which electrically connects it to two integrated electrodes; and (2) effective electrical coupling of the capillary to the system, enabling it to function as a counter electrode. Finally, the measured delay time between a pressure change in the ESEM chamber and the corresponding MS response demonstrates the feasibility of future time-resolved correlation of reaction-product formation with catalyst dynamics.

Taken together, these validations establish the LPMS platform as a promising approach for future correlative studies combining atomic-resolution ETEM imaging with localized reaction-product detection under reactive conditions. In addition, the demonstrated control over both the number and crystallographic orientation of catalyst nanoplates at the reaction site provides a promising route toward future quantitative structure–reactivity correlation under reactive conditions.

\section*{Acknowledgments}
This work was financially supported by the DFG under Project Number 544637349 and by the SFB 1633, project C04 (Grant No. 510228793). Use of equipment at the Collaborative Laboratory and User Facility for Electron Microscopy (CLUE), University of Göttingen, is gratefully acknowledged.

\section*{Competing interests}
The authors declare no competing interests.

\bibliographystyle{plainnat}

\bibliography{references}

@article{sarmah2023sustainable,
  title={Sustainable hydrogen generation and storage--a review},
  author={Sarmah, Mrinmoy Kumar and Singh, Tej Pratap and Kalita, Pankaj and Dewan, Anupam},
  journal={RSC advances},
  volume={13},
  number={36},
  pages={25253--25275},
  year={2023},
  publisher={Royal Society of Chemistry},
  doi={https://doi.org/10.1039/D3RA04148D}
}

@article{krishtalik1986energetics,
  title={Energetics of multielectron reactions. Photosynthetic oxygen evolution},
  author={Krishtalik, Lev I},
  journal={Biochimica et Biophysica Acta (BBA)-Bioenergetics},
  volume={849},
  number={1},
  pages={162--171},
  year={1986},
  publisher={Elsevier},
  doi={https://doi.org/10.1016/0005-2728(86)90107-6}
}

@article{rossmeisl2007electrolysis,
  title={Electrolysis of water on oxide surfaces},
  author={Rossmeisl, Jan and Qu, Z-W and Zhu, H and Kroes, G-J and N{\o}rskov, Jens Kehlet},
  journal={Journal of Electroanalytical Chemistry},
  volume={607},
  number={1-2},
  pages={83--89},
  year={2007},
  publisher={Elsevier},
  doi={https://doi.org/10.1016/j.jelechem.2006.11.008}
}

@article{man2011universality,
  title={Universality in oxygen evolution electrocatalysis on oxide surfaces},
  author={Man, Isabela C and Su, Hai-Yan and Calle-Vallejo, Federico and Hansen, Heine A and Mart{\'\i}nez, Jos{\'e} I and Inoglu, Nilay G and Kitchin, John and Jaramillo, Thomas F and N{\o}rskov, Jens K and Rossmeisl, Jan},
  journal={ChemCatChem},
  volume={3},
  number={7},
  pages={1159--1165},
  year={2011},
  publisher={Wiley Online Library},
  doi={https://doi.org/10.1002/cctc.201000397}
}

@article{conway1964electrochemical,
  title={Electrochemical reaction orders: Applications to the hydrogen-and oxygen-evolution reactions},
  author={Conway, BE and Salomon, M},
  journal={Electrochimica Acta},
  volume={9},
  number={12},
  pages={1599--1615},
  year={1964},
  publisher={Elsevier},
  doi={https://doi.org/10.1016/0013-4686(64)80088-8}
}

@article{song2019unconventional,
  title={An unconventional iron nickel catalyst for the oxygen evolution reaction},
  author={Song, Fang and Busch, Michael M and Lassalle-Kaiser, Benedikt and Hsu, Chia-Shuo and Petkucheva, Elitsa and Bensimon, Micha{\"e}l and Chen, Hao Ming and Corminboeuf, Clemence and Hu, Xile},
  journal={ACS central science},
  volume={5},
  number={3},
  pages={558--568},
  year={2019},
  publisher={ACS Publications},
  doi={https://doi.org/10.1021/acscentsci.9b00053}
}

@article{gono2020oxygen,
  title={Oxygen evolution reaction: Bifunctional mechanism breaking the linear scaling relationship},
  author={Gono, Patrick and Pasquarello, Alfredo},
  journal={The Journal of chemical physics},
  volume={152},
  number={10},
  year={2020},
  publisher={AIP Publishing},
  doi={https://doi.org/10.1063/1.5143235}
}

@article{hwang2020situ,
  title={In situ transmission electron microscopy on energy-related catalysis},
  author={Hwang, Sooyeon and Chen, Xiaobo and Zhou, Guangwen and Su, Dong},
  journal={Advanced Energy Materials},
  volume={10},
  number={11},
  pages={1902105},
  year={2020},
  publisher={Wiley Online Library},
  doi={https://doi.org/10.1002/aenm.201902105}
}

@article{serra2021nanoscale,
  title={Nanoscale chemical and structural analysis during in situ scanning/transmission electron microscopy in liquids},
  author={Serra-Maia, Rui and Kumar, Pawan and Meng, Andrew C and Foucher, Alexandre C and Kang, Yijin and Karki, Khim and Jariwala, Deep and Stach, Eric A},
  journal={ACS nano},
  volume={15},
  number={6},
  pages={10228--10240},
  year={2021},
  publisher={ACS Publications},
  doi={https://doi.org/10.1021/acsnano.1c02340}
}

@article{yuk2012high,
  title={High-resolution EM of colloidal nanocrystal growth using graphene liquid cells},
  author={Yuk, Jong Min and Park, Jungwon and Ercius, Peter and Kim, Kwanpyo and Hellebusch, Daniel J and Crommie, Michael F and Lee, Jeong Yong and Zettl, A and Alivisatos, A Paul},
  journal={Science},
  volume={336},
  number={6077},
  pages={61--64},
  year={2012},
  publisher={American Association for the Advancement of Science},
  doi={https://doi.org/10.1126/science.1217654}
}

@article{heide1962electron,
  title={Electron microscopic observation of specimens under controlled gas pressure},
  author={Heide, Hans Gunther},
  journal={The Journal of cell biology},
  volume={13},
  number={1},
  pages={147--152},
  year={1962},
  publisher={Rockefeller University Press},
  doi={https://doi.org/10.1083/jcb.13.1.147}
}

@article{creemer2008atomic,
  title={Atomic-scale electron microscopy at ambient pressure},
  author={Creemer, JF and Helveg, S and Hoveling, GH and Ullmann, S and Molenbroek, AM and Sarro, PM and Zandbergen, HW},
  journal={Ultramicroscopy},
  volume={108},
  number={9},
  pages={993--998},
  year={2008},
  publisher={Elsevier},
  doi={https://doi.org/10.1016/j.ultramic.2008.04.014}
}

@article{creemer2010mems,
  author={Creemer, J. Fredrik and Helveg, Stig and Kooyman, Patricia J. and Molenbroek, Alfons M. and Zandbergen, Henny W. and Sarro, Pasqualina M.},
  journal={Journal of Microelectromechanical Systems}, 
  title={A MEMS Reactor for Atomic-Scale Microscopy of Nanomaterials Under Industrially Relevant Conditions}, 
  year={2010},
  volume={19},
  number={2},
  pages={254-264},
  doi={10.1109/JMEMS.2010.2041190}}

@article{boyes1997environmental,
  title={Environmental high resolution electron microscopy and applications to chemical science},
  author={Boyes, ED and Gai, PL},
  journal={Ultramicroscopy},
  volume={67},
  number={1-4},
  pages={219--232},
  year={1997},
  publisher={Elsevier},
  doi={https://doi.org/10.1016/S0304-3991(96)00099-X}
}

@article{hansen2001atomic,
  title={Atomic-resolution in situ transmission electron microscopy of a promoter of a heterogeneous catalyst},
  author={Hansen, Thomas W and Wagner, Jakob B and Hansen, Poul L and Dahl, S{\o}ren and Tops{\o}e, Haldor and Jacobsen, Claus JH},
  journal={science},
  volume={294},
  number={5546},
  pages={1508--1510},
  year={2001},
  publisher={American Association for the Advancement of Science},
  doi={https://doi.org/10.1126/science.1064399}
}

@article{helveg2004atomic,
  title={Atomic-scale imaging of carbon nanofibre growth},
  author={Helveg, Stig and L{\'o}pez-Cartes, Carlos and Sehested, Jens and Hansen, Poul L and Clausen, Bjerne S and Rostrup-Nielsen, Jens R and Abild-Pedersen, Frank and N{\o}rskov, Jens K},
  journal={Nature},
  volume={427},
  number={6973},
  pages={426--429},
  year={2004},
  publisher={Nature Publishing Group UK London},
  doi={https://doi.org/10.1038/nature02278}
}

@article{ronge2021atom,
  title={Atom surface dynamics of manganese oxide under oxygen evolution reaction-like conditions studied by in situ environmental transmission electron microscopy},
  author={Ronge, Emanuel and Lindner, Jonas and Ross, Ulrich and Melder, Jens and Ohms, Jonas and Roddatis, Vladimir and Kurz, Philipp and Jooss, Christian},
  journal={The Journal of Physical Chemistry C},
  volume={125},
  number={9},
  pages={5037--5047},
  year={2021},
  publisher={ACS Publications},
doi={https://doi.org/10.1021/acs.jpcc.0c09806}
}

@article{lole2020dynamic,
  title={Dynamic observation of manganese adatom mobility at perovskite oxide catalyst interfaces with water},
  author={Lole, Gaurav and Roddatis, Vladimir and Ross, Ulrich and Risch, Marcel and Meyer, Tobias and Rump, Lukas and Geppert, Janis and Wartner, Garlef and Bl{\"o}chl, Peter and Jooss, Christian},
  journal={Communications Materials},
  volume={1},
  number={1},
  pages={68},
  year={2020},
  publisher={Nature Publishing Group UK London},
  doi={https://doi.org/10.1038/s43246-020-00070-6}
}

@article{raabe2012situ,
  title={In situ electrochemical electron microscopy study of oxygen evolution activity of doped manganite perovskites},
  author={Raabe, Stephanie and Mierwaldt, Daniel and Ciston, Jim and Uijttewaal, Matth{\'e} and Stein, Helge and Hoffmann, J{\"o}rg and Zhu, Yimei and Bl{\"o}chl, Peter and Jooss, Christian},
  journal={Advanced Functional Materials},
  volume={22},
  number={16},
  pages={3378--3388},
  year={2012},
  publisher={Wiley Online Library},
  doi={https://doi.org/10.1002/adfm.201103173}
}

@article{muto2019environmental,
  title={Environmental high-voltage S/TEM combined with a quadrupole mass spectrometer for concurrent in situ structural characterization and detection of product gas molecules associated with chemical reactions},
  author={Muto, Shunsuke and Arai, Shigeo and Higuchi, Tetsuo and Orita, Koji and Ohta, Shigemasa and Tanaka, Hiromochi and Suganuma, Takuya and Ibe, Masaya and Hirata, Hirohito},
  journal={Microscopy},
  volume={68},
  number={2},
  pages={185--188},
  year={2019},
  publisher={Oxford University Press},
  doi={https://doi.org/10.1093/jmicro/dfy141}
}

@article{vendelbo2014visualization,
  title={Visualization of oscillatory behaviour of Pt nanoparticles catalysing CO oxidation},
  author={Vendelbo, SB al and Elkj{\ae}r, Christian Fink and Falsig, H and Puspitasari, I and Dona, P and Mele, L and Morana, B and Nelissen, BJ and Van Rijn, R and Creemer, JF and others},
  journal={Nature materials},
  volume={13},
  number={9},
  pages={884--890},
  year={2014},
  publisher={Nature Publishing Group UK London},
  doi={https://doi.org/10.1038/nmat4033}
}

@article{plodinec2020versatile,
  title={Versatile homebuilt gas feed and analysis system for operando TEM of catalysts at work},
  author={Plodinec, Milivoj and Nerl, Hannah C and Farra, Ramzi and Willinger, Marc G and Stotz, Eugen and Schl{\"o}gl, Robert and Lunkenbein, Thomas},
  journal={Microscopy and Microanalysis},
  volume={26},
  number={2},
  pages={220--228},
  year={2020},
  publisher={Oxford University Press},
  doi={https://doi.org/10.1017/S143192762000015X}
}

@article{LINDNER2023113629,
title = {Langmuir analysis of electron beam induced plasma in environmental TEM},
journal = {Ultramicroscopy},
volume = {243},
pages = {113629},
year = {2023},
issn = {0304-3991},
doi = {https://doi.org/10.1016/j.ultramic.2022.113629},
author = {J. Lindner and U. Ross and V. Roddatis and Ch. Jooss}
}

@article{mildner2015environmental,
  title={Environmental TEM study of electron beam induced electrochemistry of Pr0. 64Ca0. 36MnO3 catalysts for oxygen evolution},
  author={Mildner, Stephanie and Beleggia, Marco and Mierwaldt, Daniel and Hansen, Thomas W and Wagner, Jakob B and Yazdi, Sadegh and Kasama, Takeshi and Ciston, Jim and Zhu, Yimei and Jooss, Christian},
  journal={The Journal of Physical Chemistry C},
  volume={119},
  number={10},
  pages={5301--5310},
  year={2015},
  publisher={ACS Publications},
  doi={https://doi.org/10.1021/jp511628c}
}

@article{jinschek2012image,
  title={Image resolution and sensitivity in an environmental transmission electron microscope},
  author={Jinschek, JR and Helveg, S},
  journal={Micron},
  volume={43},
  number={11},
  pages={1156--1168},
  year={2012},
  publisher={Elsevier},
  doi={https://doi.org/10.1016/j.micron.2012.01.006}
}

@article{maruo1997three,
  title={Three-dimensional microfabrication with two-photon-absorbed photopolymerization},
  author={Maruo, Shoji and Nakamura, Osamu and Kawata, Satoshi},
  journal={Optics letters},
  volume={22},
  number={2},
  pages={132--134},
  year={1997},
  publisher={Optical Society of America},
  doi={https://doi.org/10.1364/OL.22.000132}
}

@misc{dens_stream,
  author       = {{DENSsolutions}},
  title        = {},
  year         = {2026},
  url          = {https://denssolutions.com/},
  note         = {Accessed: 2026-04-07}
}

@article{saddeler2020effect,
  title={Effect of the size and shape on the electrocatalytic activity of Co3O4 nanoparticles in the oxygen evolution reaction},
  author={Saddeler, S and Hagemann, U and Schulz, S},
  journal={Inorganic chemistry},
  volume={59},
  number={14},
  pages={10013--10024},
  year={2020},
  publisher={ACS Publications},
  doi={https://doi.org/10.1021/acs.inorgchem.0c01180}
}

@misc{comsol54,
  author       = {{COMSOL AB}},
  title        = {COMSOL Multiphysics\textregistered{} v. 5.4},
  howpublished = {\url{www.comsol.com}},
  address      = {Stockholm, Sweden},
  year         = {2018}
}

@article{BIRAN2026114328,
title = {Open gas-cell transmission electron microscopy at 0.5 Å information limit},
journal = {Ultramicroscopy},
volume = {282},
pages = {114328},
year = {2026},
issn = {0304-3991},
doi = {https://doi.org/10.1016/j.ultramic.2026.114328},
author = {Idan Biran and Frederik Dam and Sophie Kargo Kaptain and Ruben Bueno Villoro and Maarten Wirix and Christian Kisielowski and Peter C.K. Vesborg and Jakob Kibsgaard and Thomas Bligaard and Christian D. Damsgaard and Joerg R. Jinschek and Stig Helveg},
}

\end{document}